\documentclass[%
 reprint,
 amsmath,amssymb,
 aps,
floatfix,
]{revtex4-2}

\usepackage{graphicx}
\usepackage{dcolumn}
\usepackage{bm}
\usepackage{physics}

\usepackage{color,soul}
\usepackage{xspace}
\usepackage{xcolor}

\begin{document}

\preprint{APS/123-QED}

 \title{{Phase Amplification in Spinodal Decomposition of Immiscible Fluids with Interconversion of Species}}

\author{Nikolay A. Shumovskyi}%
\affiliation{Department of Physics, Boston University, Boston, MA 02215, USA}

\author{Thomas J. Longo}%
\affiliation{Institute for Physical Science and Technology, University of Maryland, College Park, MD 20742, USA}

\author{Sergey V. Buldyrev}%
\email{buldyrev@yu.edu}
\affiliation{ Department of Physics, Yeshiva University, New York, NY 10033, USA    \\                  Department of Physics, Boston University, MA 02215, USA}

\author{Mikhail A. Anisimov}%
\email{anisimov@umd.edu}
\affiliation{Department of Chemical and Biomolecular Engineering and Institute for Physical Science
and Technology, University of Maryland, College Park, MD 20742, USA}

\date{\today}

\begin{abstract}
A fluid composed of two molecular species may undergo phase segregation via spinodal decomposition.  However, if the two molecular species can interconvert, \textit{e.g.} change their chirality, then a phenomenon of phase amplification, which has not been studied so far, emerges. As a result, eventually, one phase will completely eliminate the other one. We model this phenomenon on an Ising system which relaxes to equilibrium through a hybrid of Kawasaki-diffusion and Glauber-interconversion dynamics. By introducing a probability of Glauber-interconversion dynamics, we show that the particle conservation law is broken, thus resulting in phase amplification. We characterize the speed of phase amplification through scaling laws based on the probability of Glauber dynamics, system size, and  distance to the critical temperature of demixing.
\end{abstract}

\maketitle


\paragraph*{\label{sec:Intro} Competing Dynamics: The Phenomenon of Phase Amplification}-When a fluid, composed of two immiscible molecular species, is quenched at appropriate concentration from a high temperature to a temperature below the critical point of demixing, into the unstable (spinodal) region, the fluid will phase separate into two stable phases - a process known as spinodal decomposition \cite{cahn_phase_1965}. During spinodal decomposition, if the molecular species may easily interconvert (\textit{e.g.} chiral molecules \cite{latinwo_molecular_2016}, chiral crystals \cite{chiral_Debenedetti_2013}, or two isomorphs in a polyamorphic fluid \cite{anisimov_thermodynamics_2018}), then the phenomenon of phase amplification, when one phase grows at the expense of another one, can be observed. 

To model this phenomenon, we use the fact that the Ising model for an anisotropic ferromagnet and the lattice gas model for a ﬂuid are mathematically equivalent \cite{lee_statistical_1952}. Within the same universality class, systems demonstrate the same critical singularities and the same critical equation of state, provided that the appropriately defined order parameters, $\phi$, have the same symmetry. The one-component-vector order parameter ($\phi$, the magnetization) in the Ising model and the scalar order parameter ($\phi = 1-2\rho$, where $\rho$ is the density) in the lattice gas posses the same symmetry \cite{kogut_introduction_1979,senthil_z_2_2000}. In this work, we consider the incompressible symmetric binary mixture \cite{xu_phase_2004,kubo}, a popular version of the lattice gas model, where the conserved order parameter is $\phi = 1-2x$, in which $x$ is the fraction of cells occupied by one of the components. 

However,  while the binary lattice and Ising model are equivalent in thermodynamics, they are fundamentally different in dynamics. The order parameter, the fraction of occupied cells, associated with fluid phase separation is conserved, while magnetization is not conserved. Thus, the order parameters in the binary lattice and Ising model belong to different universality classes in dynamics \cite{hohenberg_theory_1977}. This difference is characterized by the dynamic critical exponent $z =2$ (Ising-model dynamics) and $z = 4$ (binary-lattice dynamics in the mean-field approximation) \cite{hohenberg_theory_1977,Static_Das_2006}. 

\begin{figure}
    \includegraphics[width=0.49\linewidth]{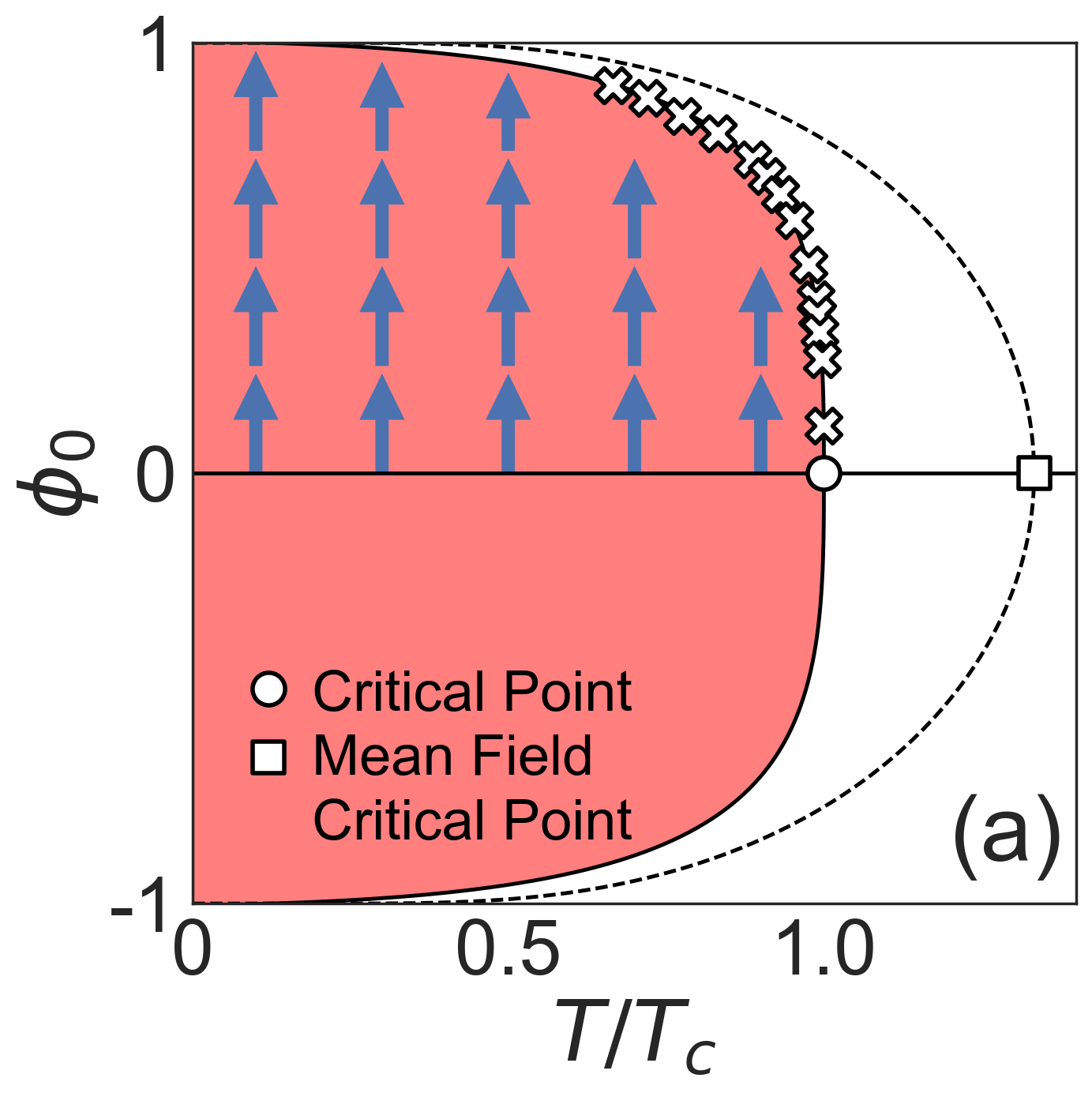}
    \includegraphics[width=0.49\linewidth]{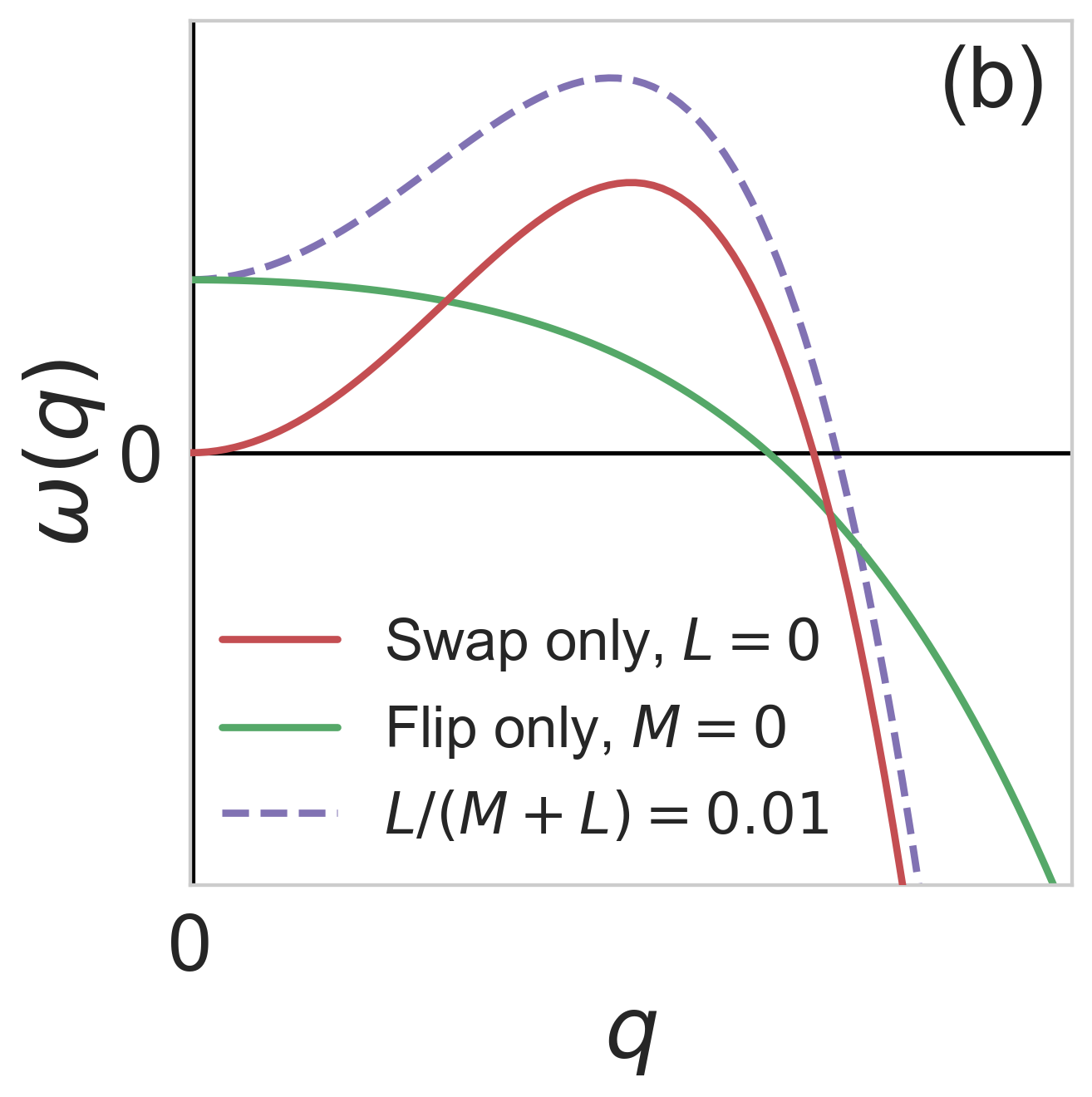}
    \caption{(a) The spontaneous equilibrium order parameter ($\phi=\phi_0$) in the lattice-gas / binary-lattice mixture along the liquid-vapor phase coexistence (red domain). One of the two alternative magnetizations ($\phi_0 > 0$ and $\phi_0 < 0$) in the Ising ferromagnet in zero field is shown in the red domain with blue arrows. The solid curve is the crossover from mean-ﬁeld behavior (dashed) to the asymptotic scaling power law $\phi \propto \Delta \hat{T}^\beta$ with $\beta=0.326$ \cite{povodyrev_crossover_1999,kim_crossover_2003}, while the crosses are our simulation data {for $\ell = 100$ and averaged over 1,000 realizations}. (b) The growth rate solution to Eq.~(\ref{Eqn-R(q)}), known as the amplification factor, for three probabilities: $p_r=0$ (pure swapping dynamics - $L = 0$), $p_r=1$ (pure flipping dynamics - $M = 0$), and $p_r = L/(M+L) = 0.01$ (mixed dynamics) after quenching the system into the unstable region at $\Delta \hat{T} = 0.1$.}
    \label{Fig-1}
\end{figure}

To illustrate this difference, consider a binary lattice of particles. In the absence of an external ordering field, but in the presence of fluctuations of density, this system will remain in equilibrium provided that the particles can arrange themselves to minimize the free energy. To do this, the particles will ``swap'' locations with one another until the energy is minimized. As a result, below the critical temperature of demixing, two equilibrium fluid phases must coexist to conserve the total number of particles (occupied cells). Juxtapose this with the same lattice of up and down spins in the Ising model. Since the positions of the magnetic dipoles are fixed, to approach equilibrium, the spins will ``flip'' to minimize the free energy. Thus, in the Ising ferromagnet, only one of the alternative magnetizations, positive or negative, will survive - see Fig.~\ref{Fig-1}a. Since the interface between the two alternative magnetic phases is energetically costly, eventually, one magnetic domain will win over the other.

Therefore, a phase separating fluid with interconversion of species can be described through a hybrid model combining diffusion-swapping dynamics, where the total number of particles of each species is conserved, and interconversion-flipping dynamics, where the total number of particles of each species is not conserved. By introducing a probability of interconversion, the conservation of the number of particles for a specific type will eventually be broken and the striking phenomenon of phase amplification (or “phase bullying”, as originally coined by Latinwo, Stillinger, and Debenedetti \cite{latinwo_molecular_2016}) can be observed. 

In order to clarify the physics of phase amplification, we consider a simple Ising system that utilizes a hybrid model combining swapping (conserved order-parameter) and flipping (nonconserved order-parameter) dynamics distinguished through a certain probability for the order parameter to exhibit interconversion dynamics. The characteristic time evolution equation for the growth of the order parameter, which contains both nonconserved and conserved features, is given by
\begin{equation}\label{Eqn-R(q)}
    \pdv{\phi}{t} = -L\mu + M\nabla^2\mu
\end{equation}
\cite{cahn_phase_1965,cahn_microscopic_1977,li_non-equilibrium_2020,MFT_PT_2021} where the exchange chemical potential, $\mu$, is the deviation of the chemical-potential difference between the two species from its equilibrium value - $\mu = 0$. It spatially depends on the order parameter (such that it characterizes local inhomogeneities within the system), the reduced distance to the critical temperature $\Delta \hat{T} = (T_\text{c}-T)/T_\text{c}$, and the correlation length of fluctuations, $\xi$. The first term in Eq.~(\ref{Eqn-R(q)}) describes the relaxation of the nonconserved order-parameter dynamics to equilibrium, and the second term describes the relaxation of the conserved order-parameter dynamics to equilibrium. The kinetic Onsager coefficients $L$ and $M$ correspond to the flipping and swapping dynamics respectively. When $M=0$, the order parameter grows according to pure flipping dynamics, while when $L=0$, the order parameter grows according to Cahn-Hilliard theory of spinodal decomposition \cite{cahn_phase_1965,cahn_microscopic_1977} (pure swapping dynamics). We define the probability that the system will exhibit interconversion of species as $p_r = L/(M + L)$. The solution to Eq.~(\ref{Eqn-R(q)}) which characterizes the rate of spinodal decomposition, $\omega(q)$, is known as the ``amplification factor'' \cite{cahn_phase_1965,MFT_PT_2021}, and it is illustrated in Fig.~\ref{Fig-1}b for three different probabilities. If $p_r=1$, then, as the system relaxes to equilibrium, eventually one phase will completely eliminate the other one via phase amplification. If $p_r = 0$, the system will reach phase coexistence at equilibrium (no phase amplification), and if $0 < p_r < 1$, we show in this work that the rate of phase amplification depends on $p_r$, the system size $\ell$, and the distance to the critical temperature $\Delta \hat{T}$.

The domain growth \cite{mazenko_theory_1984} and symmetry breaking in chiral molecules \cite{latinwo_molecular_2016} and in chiral crystals \cite{chiral_Debenedetti_2013} have already been studied. However, phase amplification, the phenomenon of the growth of a nonconserved order parameter, at different interconversion probabilities has not yet been addressed. In this Letter, we report on the results of Monte Carlo (MC) simulations of a simple Ising model that utilizes a hybrid combination of Kawasaki-swapping \cite{kawasaki_diffusion_1966} and Glauber-flipping \cite{glauber_timedependent_1963} dynamics to account for the conserved and nonconserved dynamics of the order parameter respectively. Previous computational studies of this combination have been studied by Glotzer \textit{et al.} \cite{glotzer_monte_1994,glotzer_reaction-controlled_1995}, but their model has been shown not to be applicable to equilibrium systems \cite{lefever_comment_1995,carati_chemical_1997,lamorgese_spinodal_2016}. In this work, we present an equilibrium formulation of the model proposed by Glotzer \textit{et al.} \cite{MFT_PT_2021}, and we show that under equilibrium conditions, the competition of swapping (Kawasaki) and flipping (Glauber) dynamics produces the phenomenon of phase amplification. We characterize the rate of phase amplification through the probability of Glauber dynamics, system size, and distance to the critical temperature. We also provide scaling arguments for the topological behavior of the resultant structures that occur during phase amplification.

\paragraph*{Model Description} - We perform mixed Kawasaki-Glauber dynamics on an Ising-spin system in zero field arranged on an $\ell \times \ell \times \ell$ cubic lattice using the conventional Ising model Hamiltonian \cite{kubo,Huang}
\begin{equation}
    H = -\frac{\epsilon}{2}\sum_{i=1}^{\ell^3}\sum_{j\in\Omega(i)}s_i s_j
\end{equation}
where $s_i$, $s_j = \pm 1$ are spins, $\Omega(i)$ is the set of 6 nearest neighbors of spin $i$, and $\epsilon$ is interaction energy. The critical temperature of this system is $T_{\text{c}}=4.5115(1)\epsilon/k_B$ \cite{Heuer93}, where $k_B$ is Boltzmann's constant. We start simulations with a random spin configuration in which $\ell^3/2$ spins are positive and $\ell^3/2$ spins are negative. In addition, we assume that at each MC step the probability of a random spin flip (a Glauber step) is $p_r$, while the probability of swapping a randomly selected pair of nearest neighbor spins (a Kawasaki step) is $1-p_r$. Each step is accepted with a standard Metropolis criterion \cite{metropolis_basic_1963}. We introduce a size-independent MC time as $t=n/\ell^3$, where $n$ is the total number of MC steps. The frequency of spin flipping is absorbed into the time step, $\delta t$, so the Onsager coefficients, $L$ and $M$ (and consequently $p_r$), do not depend on temperature.

\begin{figure}
    \centering
    \includegraphics[width=0.49\linewidth]{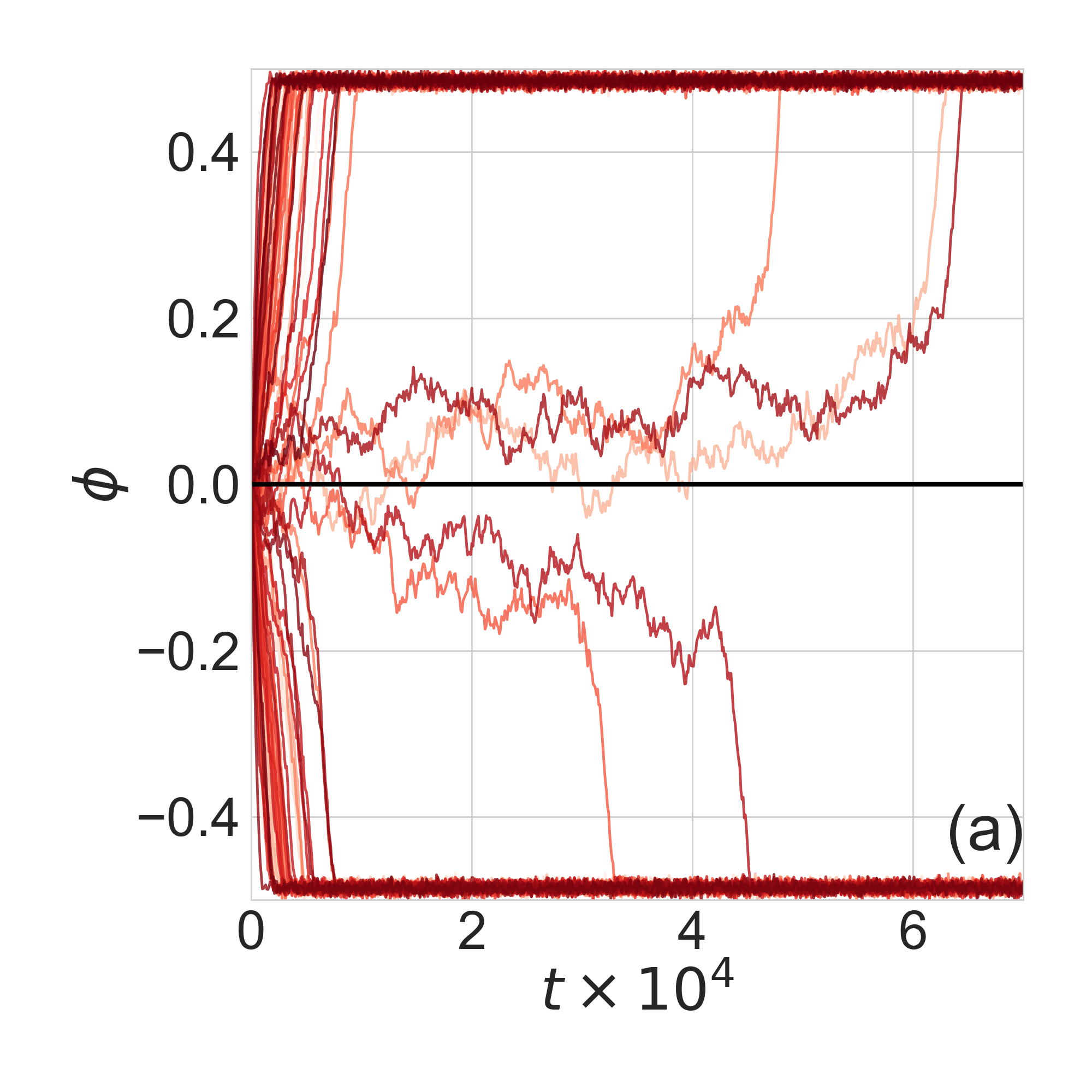}
    \includegraphics[width=0.49\linewidth]{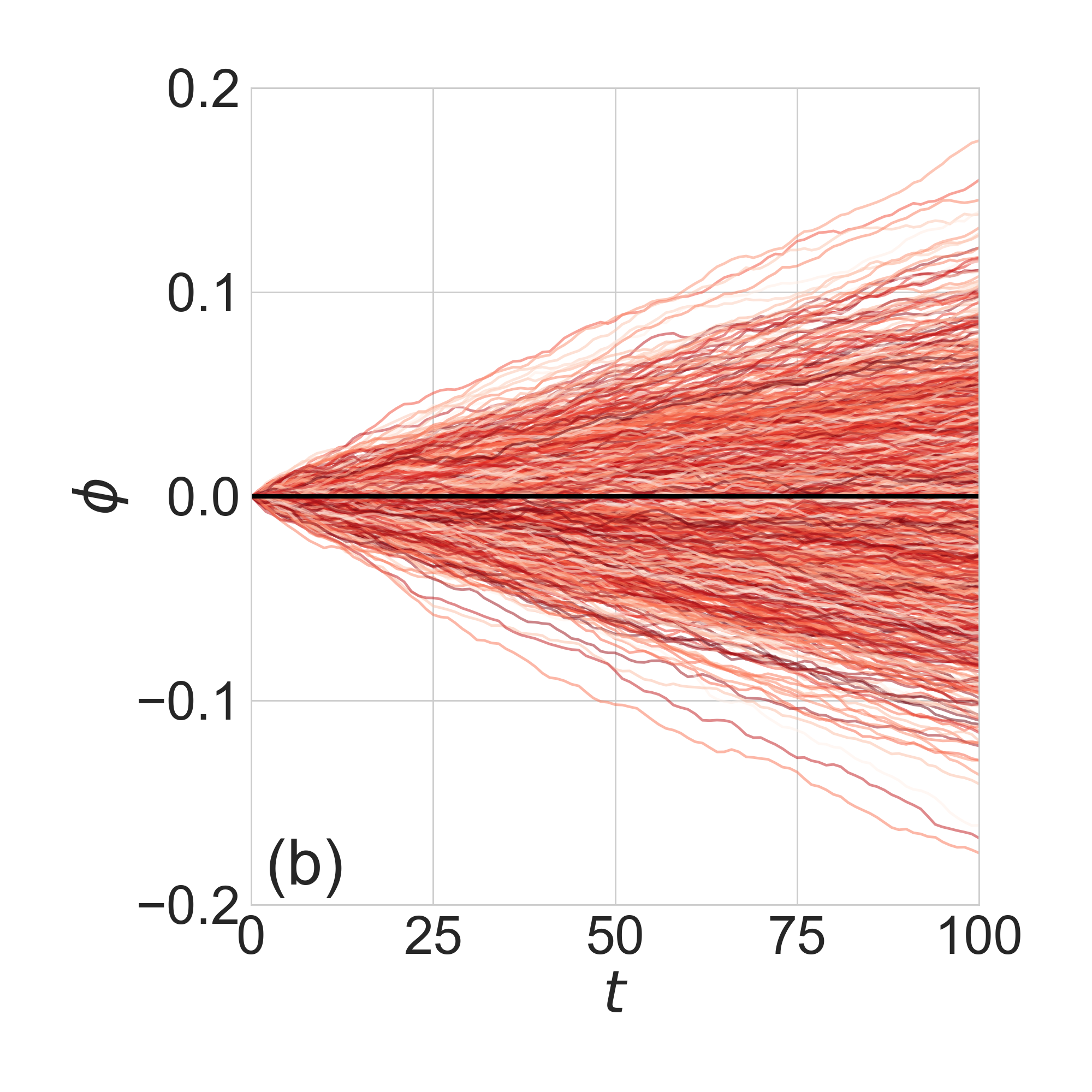}
    \includegraphics[width=0.49\linewidth]{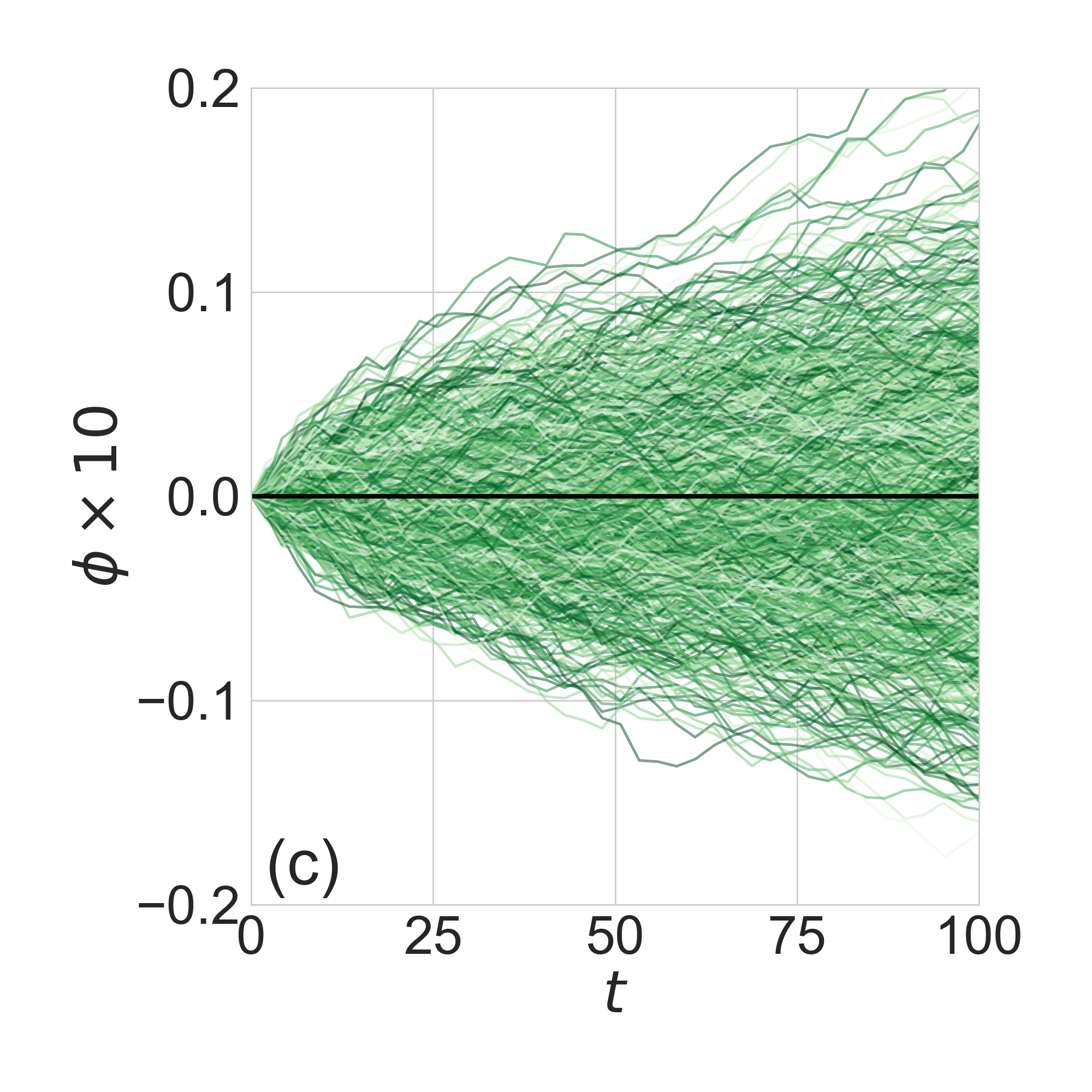}
    \includegraphics[width=0.49\linewidth]{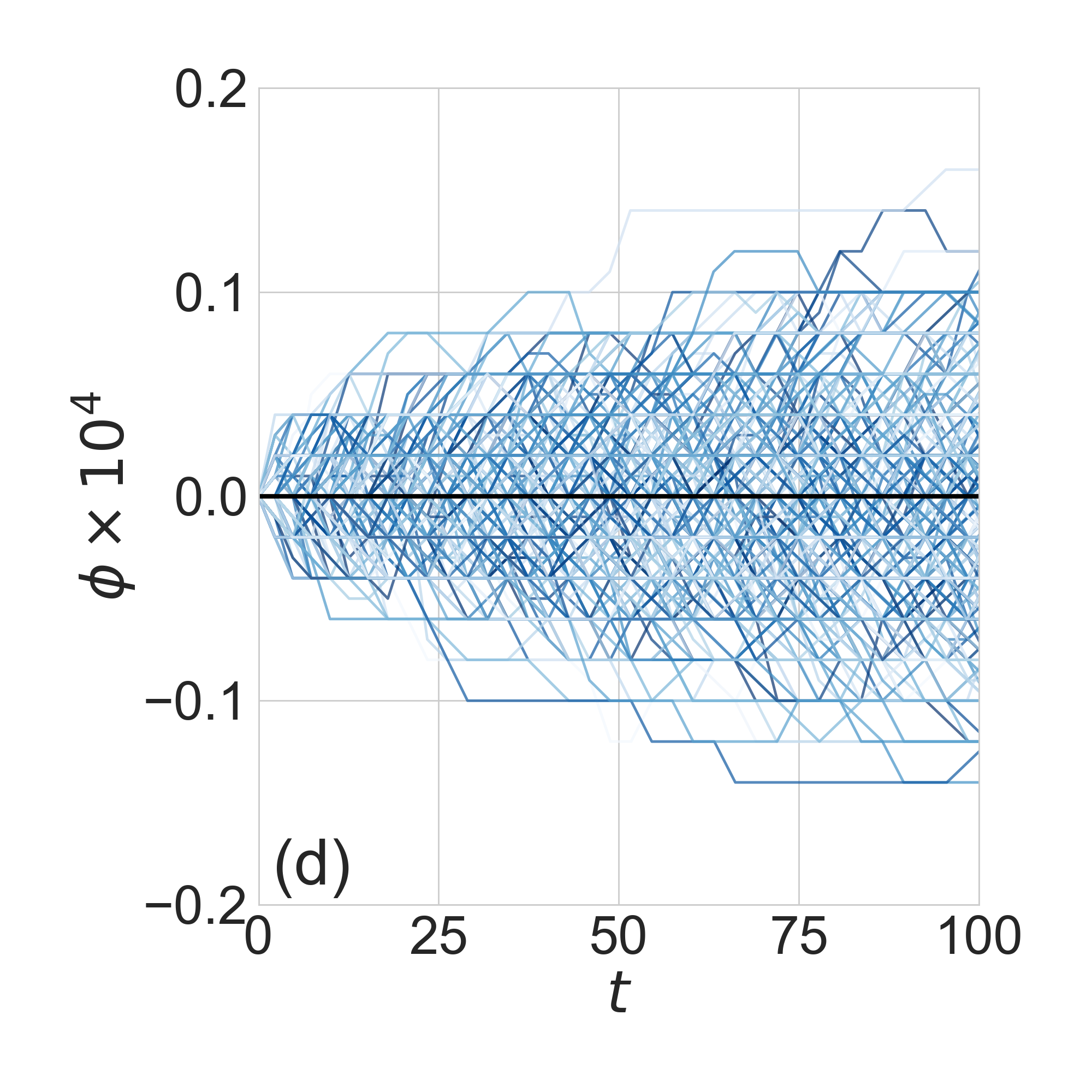}
    \caption{Phase amplification - the growth of the order parameter for different probabilities of Glauber dynamics at a system size $\ell = 100$. (a) Full-time behavior, $T=4.4$ and 100 realizations, $p_r = 1$; (b-d) initial time behavior, $T=4.0$ and 1,000 realizations: (b) $p_r=1.0$, (c) $p_r=0.1$, and (d) $p_r=1.0\times 10^{-7}$. The solid horizontal line, $\phi = 0$, corresponds to Kawasaki dynamics, $p_r=0$.}
    \label{Fig-BullyingProb}
\end{figure}

\paragraph*{Results} - In zero field, the order parameter for the Ising system generates phase domains when $T < T_{\text{c}}$. These domains will either grow or collapse according to the time evolution equation, Eq.~(\ref{Eqn-R(q)}) - see Fig.~\ref{Fig-1}b, such that the competition between the two types of dynamics determines the initial growth of these domains. For a hybrid model of Glauber and Kawasaki dynamics, this competition eventually results in one phase completely eliminating the other - see Fig.~\ref{Fig-BullyingProb}a. However, the rate of phase amplification depends on the probability for the system to follow Glauber (nonconserved) order-parameter dynamics, see Fig.~\ref{Fig-BullyingProb}(b-d) where the phenomenon of phase amplification is shown for different interconversion probabilities from pure Glauber, $p_r=1$, to extremely low probability $p_r = 1.0\times 10^{-7}$. For this extremely low probability, most realizations just fluctuate around the average value of the order parameter, $\langle\left|\phi\right|\rangle=0$. In principle, any nonconserved order-parameter dynamics, even with an extremely small probability, may eventually lead to phase amplification, but this is only true for an extremely large computational (or observational) time. For a certain low probability, these conditions would not be achievable in any computational or experimental studies.

\begin{figure}
    \centering
    \includegraphics[width=0.49\linewidth]{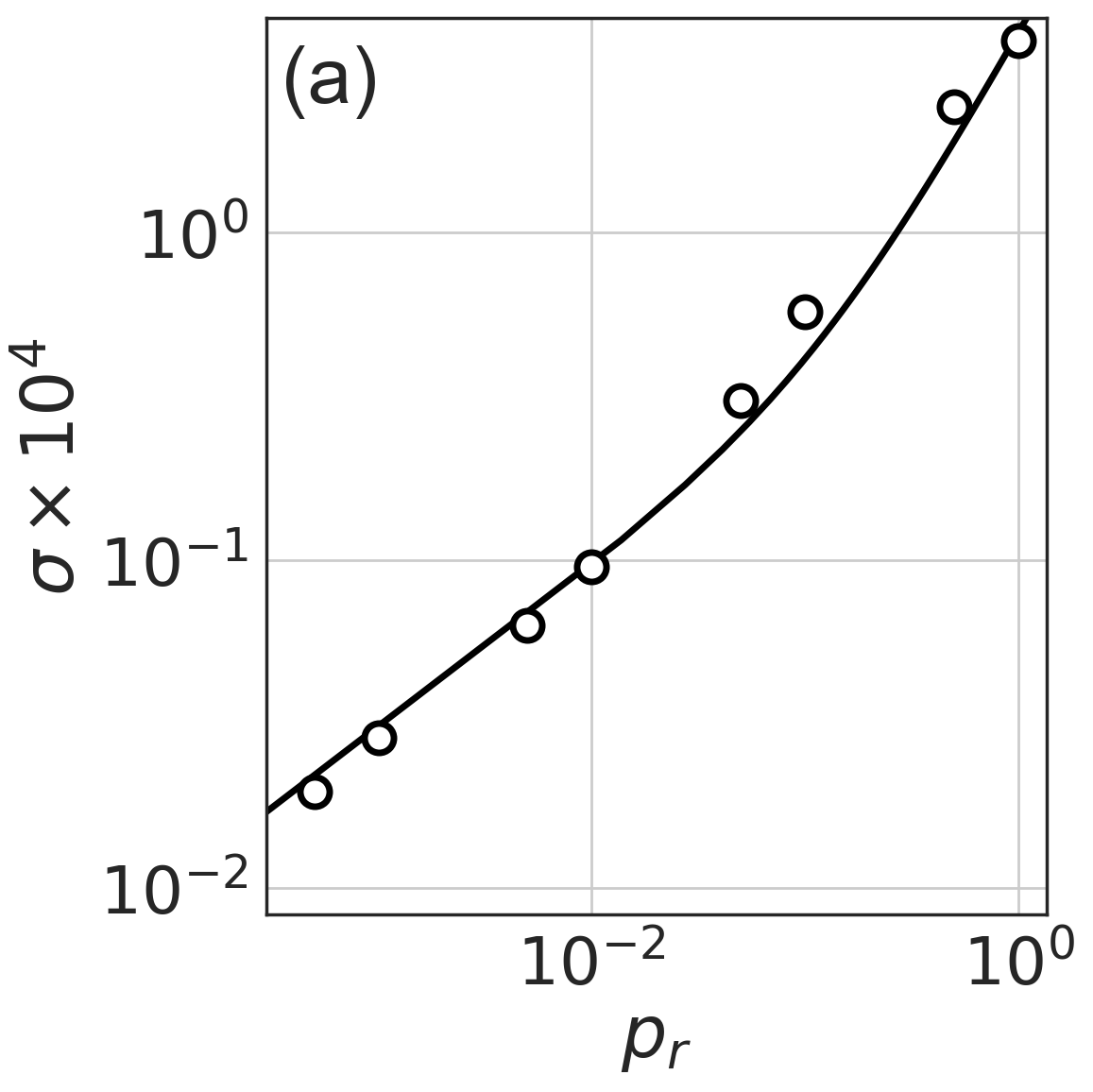}
    \includegraphics[width=0.49\linewidth]{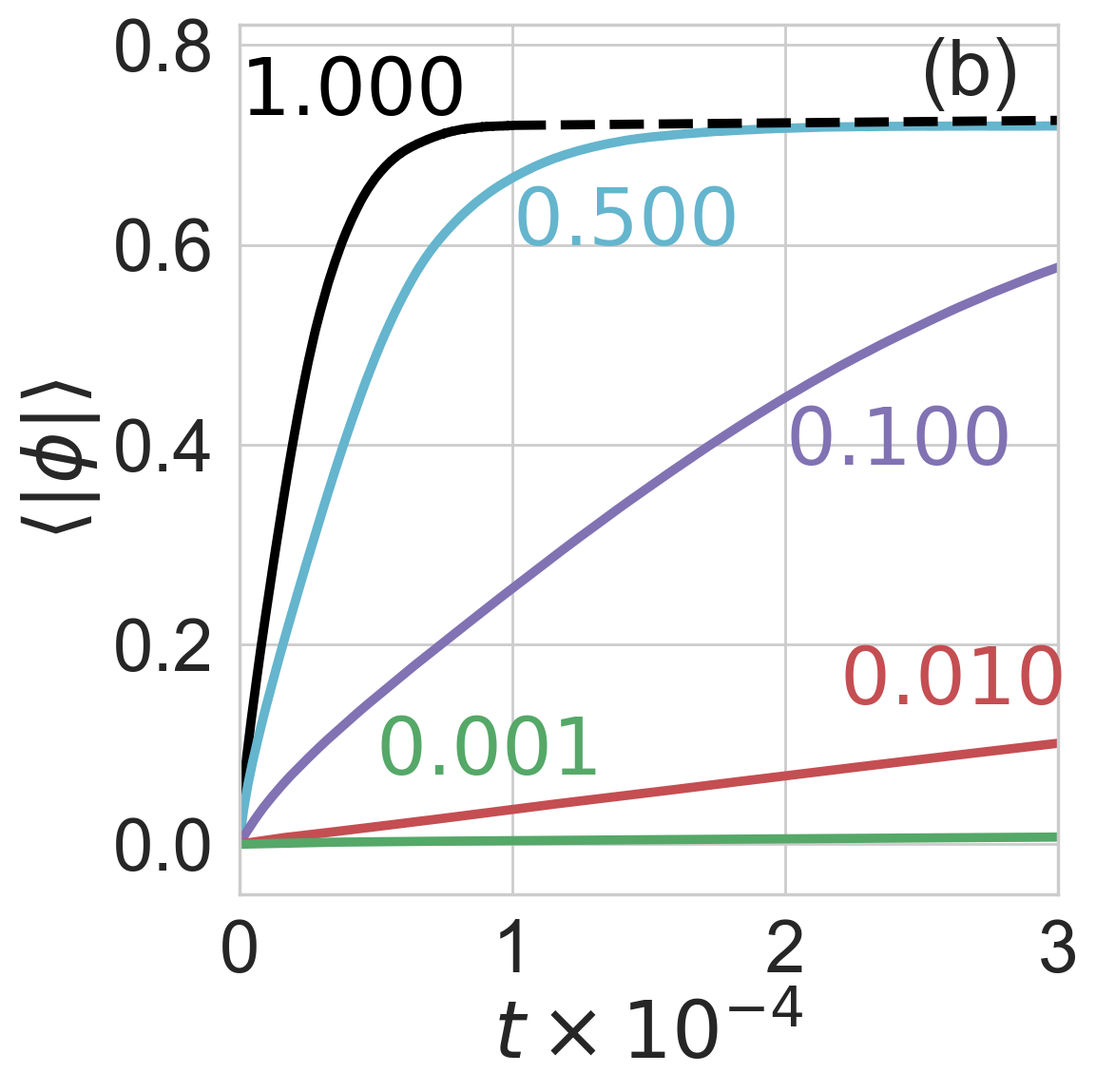}
    \includegraphics[width=0.49\linewidth]{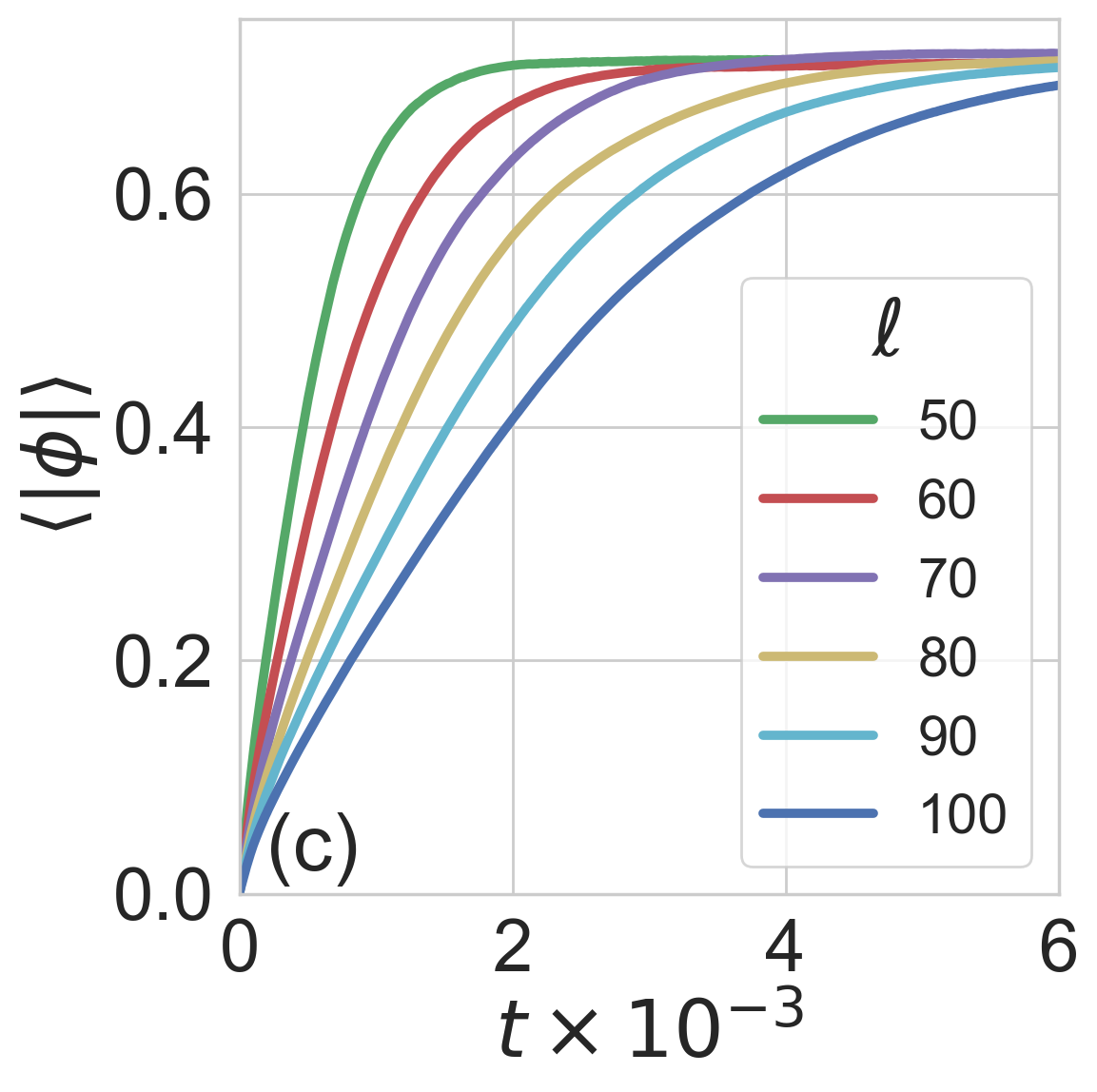}
    \includegraphics[width=0.49\linewidth]{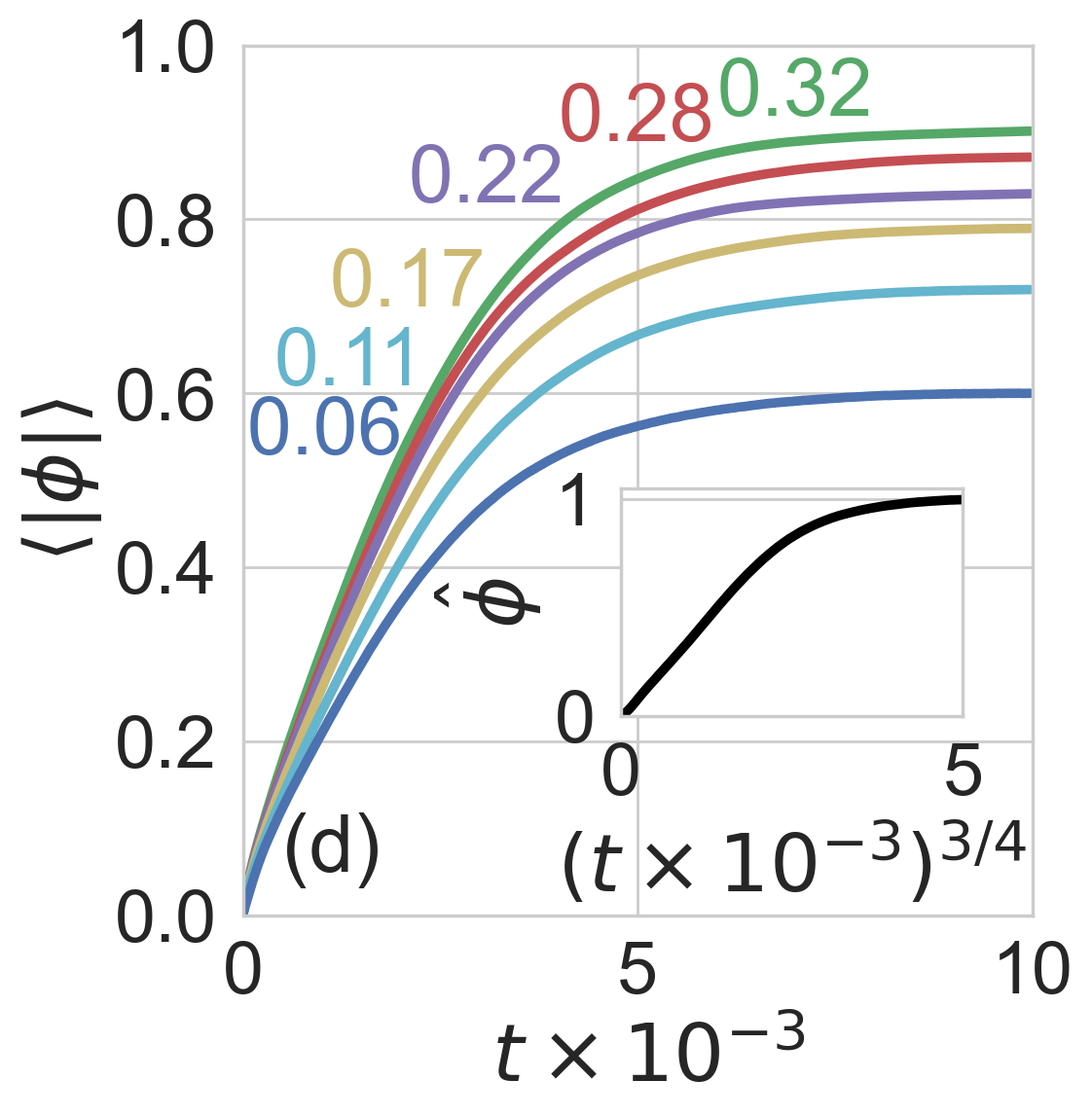}
    \caption{The evolution of the order parameter during phase amplification. (a) The RMS of the distribution of the growth rates for different probabilities captured at the same time, $t=300$. The solid curve is the crossover between $\sigma \propto \sqrt{p_r}$ and $\sigma \propto p_r$, approximated as $\sigma = a\sqrt{p_r}(1+bp_r)/(1+\sqrt{p_r})$. (b-d) The growth of the order parameter for different (b) probabilities at $\Delta \hat{T} = 0.11$ and $\ell=100$, (c) system sizes at $p_r = 1.0$ and $\Delta \hat{T} = 0.11$, and (d) distances to the critical point at $p_r=1.0$ and $\ell=100$; the colored values in (b,d) correspond to the colored curves of different $p_r$ and $\Delta \hat{T}$ respectively. The inset of (d) shows the power law for the initial growth of the reduced order parameter, $\phi/\phi_0 \propto t^{3/4}$.}
    \label{fig-Averages}
\end{figure}

One can notice from Fig.~\ref{Fig-BullyingProb}a, that even for systems with pure Glauber dynamics, there is a small fraction of realizations that survive for a long time, but eventually, one phase completely eliminates the other. We attribute this effect to the accidental formation of zero-curvature interfaces during the domain growth. We observed that at least two types of zero curvature interfaces are possible: a planar interface and a Schwarz-P interface \cite{schwarz_gesammelte_1972}. Of these two types, only the planar surface corresponds to stable equilibrium against phase amplification. In this case, a bump with positive curvature produced by a fluctuation shrinks, while a cavity with negative curvature will flatten. In periodic boundary conditions, this interface forms a strip with two parallel surfaces. Fluctuations will only produce random changes to the width of this strip, $w=\ell \Delta \hat{\phi}/2$ (where $\Delta \hat{\phi} = 1-\phi/\phi_0$ is the reduced deviation from the equilibrium order parameter), corresponding to the longer lasting realizations in Fig.~\ref{Fig-BullyingProb}a. Eventually, when $w$ becomes comparable to $\xi$, the domain will be punctured and a hole, with average positive curvature due to the growing domain, will be formed causing the strip to quickly disappear. In periodic boundary conditions, or locally, the growing domains can form other surfaces with zero curvature, for example, a Schwarz-P surface \cite{schwarz_gesammelte_1972}. This is especially likely when $p_r$ is small and the system has time to evolve largely according to Kawasaki dynamics. Once such an interface is formed, the amplification process is ``frozen'', and the order parameter follows random walk behavior. However, such interfaces are unstable against fluctuations and will collapse when the growing phase forms an interface with negative curvature to break the phase domain of the receding phase. 

One of the most evident characteristic of phase amplification is the increase in the width of the distribution of the growth rates as the Glauber-dynamics probability increases. Assuming that the distribution of the average rates (slopes) in Fig.~\ref{Fig-BullyingProb}{(b-d)} is Gaussian, we show in Fig.~\ref{fig-Averages}a that the standard deviation of this distribution (calculated at the same time, $t=300$) changes from $\sigma \propto \sqrt{p_r}$ to $\sigma \propto p_r$ as the system transitions from Kawasaki to Glauber dynamic behavior. Next, we study the absolute value of the order parameter, averaged over $1000$ independent realizations of the evolution, as a function of the three key parameters: {Glauber} probability $p_r$, system size $\ell$ {\cite{Footnote1}}, and distance to the critical point $\Delta \hat{T}$. As shown in Fig.~\ref{fig-Averages}(b-d), phase amplification is faster for larger $p_r$, smaller $\ell$, and further distance to the critical point (larger $\Delta \hat{T}$).

\begin{figure}
    \centering
    \includegraphics[width=0.49\linewidth]{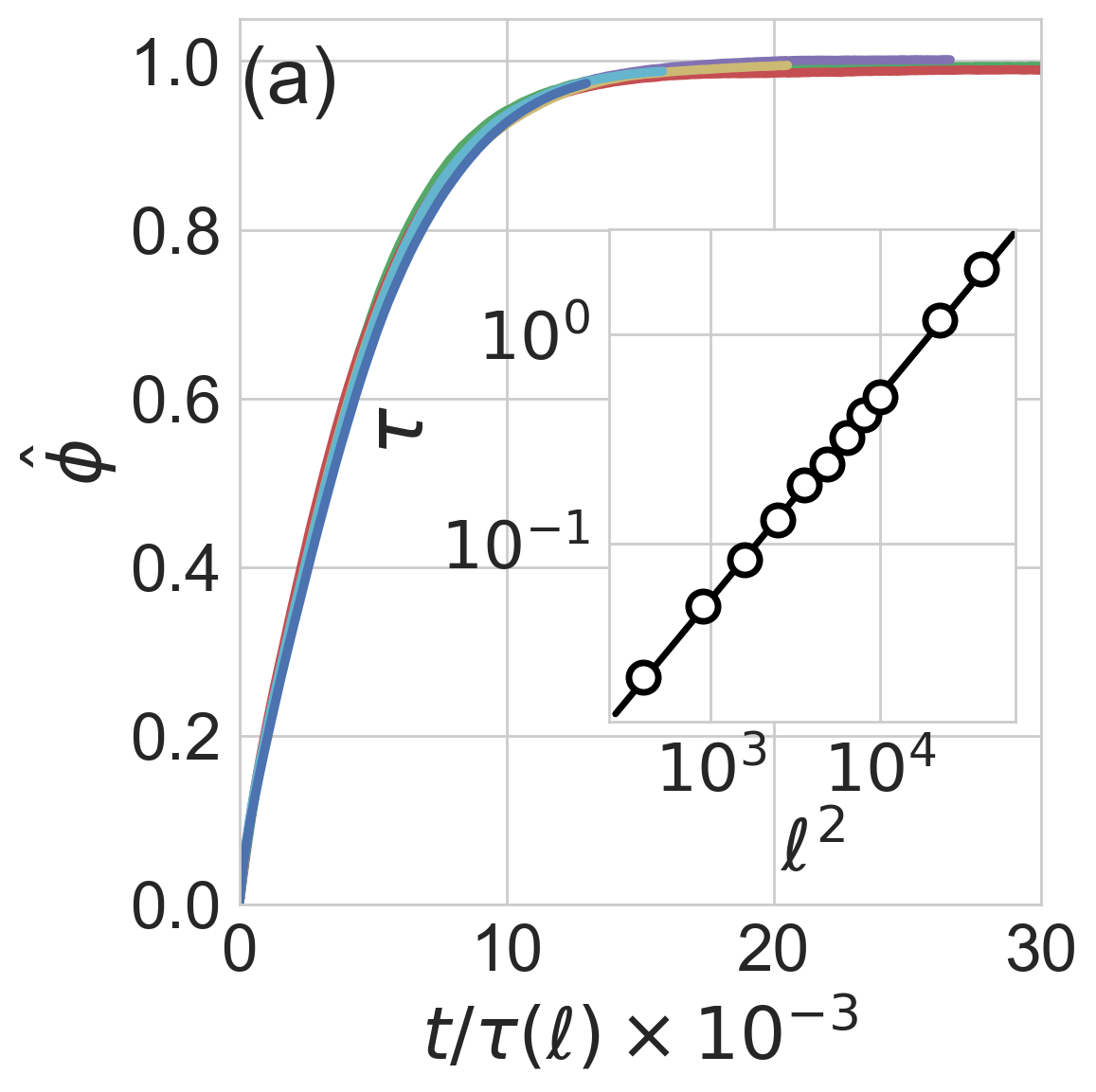}
    \includegraphics[width=0.49\linewidth]{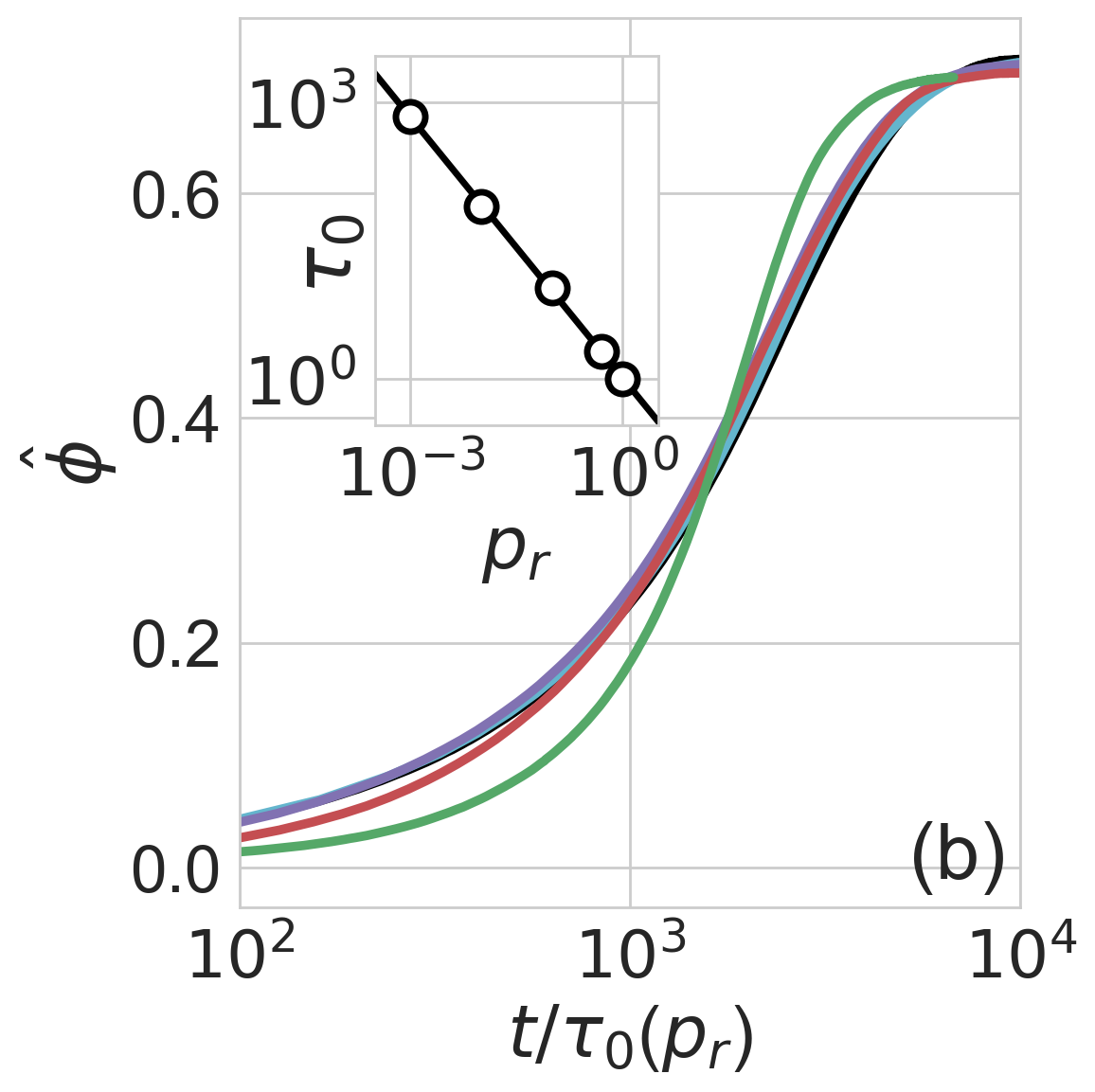}
    \caption{Scaling properties of the growth of the reduced order parameter, $\hat{\phi}=\langle\left|\phi\right|\rangle/\phi_0$. (a) The order parameter growth with time rescaled by system size. The size dependence of the rescaling parameter, $\tau(\ell)$, is shown in the inset; in the log-log scale with a slope of 1. The colors are the same as in Fig.~\ref{fig-Averages}c. (b) The order parameter growth with time rescaled by probability; the rescaling parameter $\tau_0(p_r)$, inversely proportional to the probability, is shown in the inset. The colors are the same as in Fig.~\ref{fig-Averages}b.}
    \label{fig-scaling}
\end{figure}

By reducing the order parameter by its equilibrium value ($\hat{\phi}=\langle\left|\phi\right|\rangle/\phi_0$) and rescaling the time as $t/\tau(\ell)$ the system-size and temperature dependent $\phi(t)$ are collapsed as shown in Fig.~\ref{fig-scaling}a; the characteristic time $\tau(\ell)$ is proportional to $\ell^2$ as shown in the inset for {$p_r=1$}. After introducing another characteristic scaling time $\tau(p_r)$, for relatively large probability ($p_r \geq 0.01$), we are able to collapse the order parameter growth in a set of curves which cross at the same inflection point as shown on Fig.~\ref{fig-scaling}b. As a result, for relatively large $p_r$, we may neglect the effects of Kawasaki dynamics on the system and develop scaling arguments to describe the growth of the order parameter under pure Glauber dynamics.

\begin{figure}
    \centering
    \includegraphics[width=0.49\linewidth]{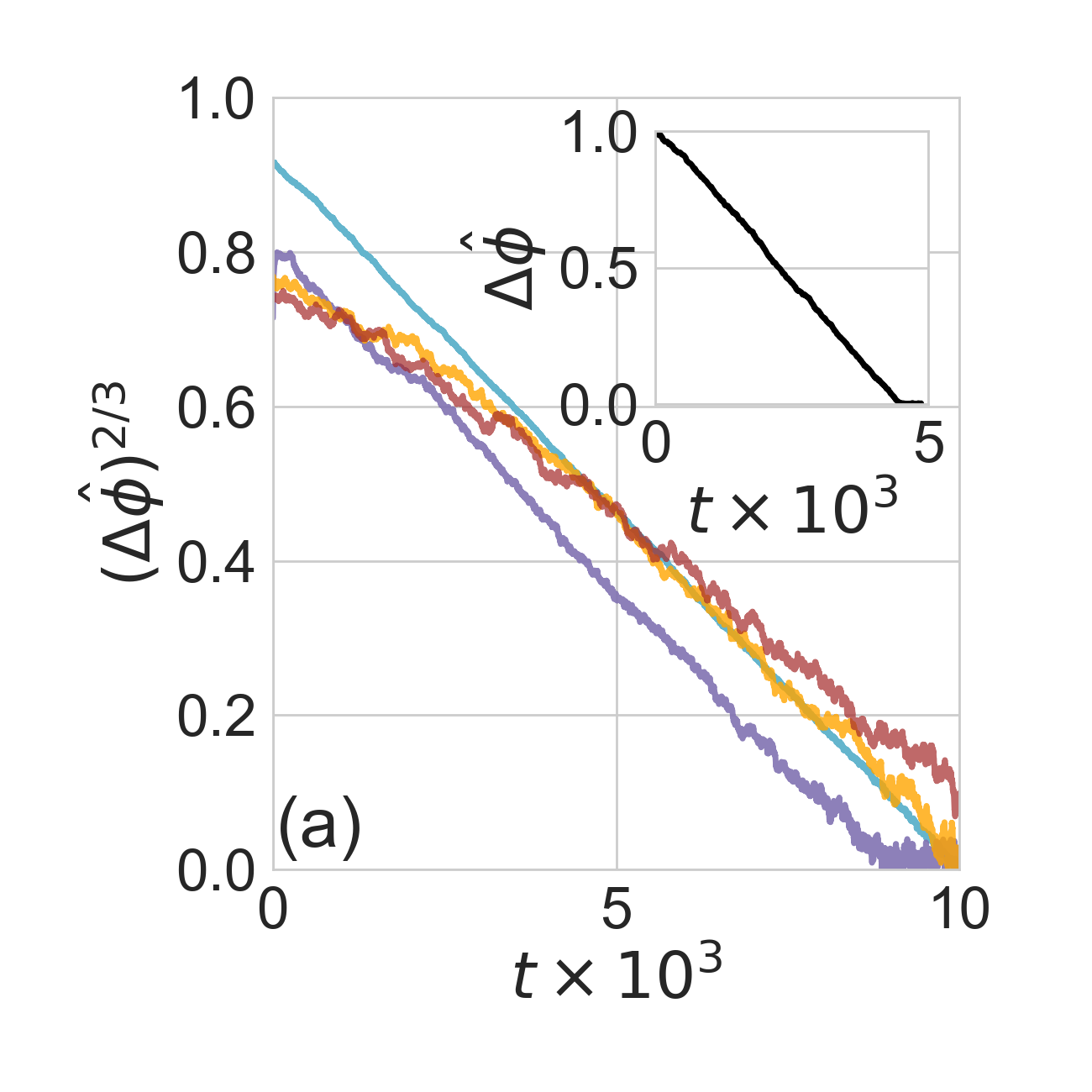}
    \includegraphics[width=0.49\linewidth]{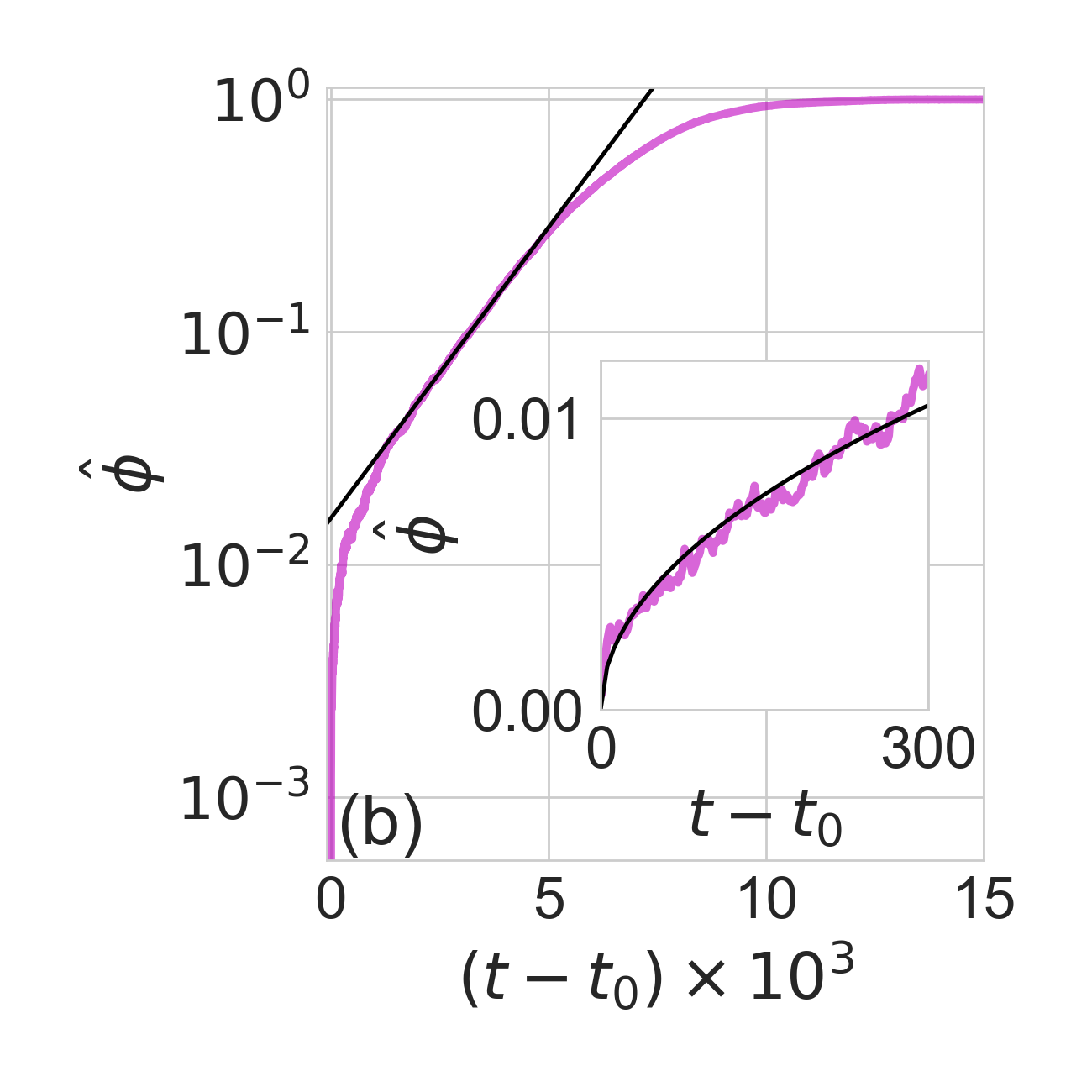}
    \caption{Topological characteristics of the time dependence of phase amplification for $\ell = 100$. (a) For spherical domains, the reduced deviation from the equilibrium order parameter, $\Delta\hat{\phi}=1-\phi/\phi_0$ scales as $t^{3/2}$. This is shown for temperatures, $\Delta \hat{T}$, as: $0.11$ (cyan), $0.025$ (purple), $0.014$ (orange), and $0.009$ (brown). The inset shows the effect for a cylindrical domain at  $\Delta \hat{T} = 0.11$. (b) A zero curvature Schwarz-P interface is initially formed by simulating a system with Kawasaki dynamics ($p_r=0$) for a long time. At $t-t_0 = 0$, the system obtains Glauber dynamics ($p_r=1$) and the collapse of one of the phases is shown. Amplification transitions from random-walk behavior, $\sqrt{t}$, at short times (see inset) to exponential behavior before saturation, shown by the straight line.}
    \label{Fig-timedep}
\end{figure}

\paragraph*{Scaling Analysis of Phase Amplification for Large $p_r$} - To develop scaling arguments for the growth of the order parameter, we have observed that phase amplification occurs when one phase domain forms an interface with negative curvature against another, such that domains with positive curvature will quickly disappear. In Fig.~\ref{Fig-timedep}a, we illustrate this process by generating a system in the initial configuration of spherical or cylindrical domains and evolve the system under pure Glauber dynamics. Our simulations show that phase amplification of these topologies occurs with rates: $\phi\propto t^{3/2}$ and $\phi \propto t$ for the sphere and cylinder respectively. This observation is a result of the Kelvin theory \cite{Huang,allen_microscopic_1979} that the radius of curvature between domains grows with time as $1/q\sim\sqrt{t}$.  Indeed, it is possible to show that the domain encapsulated by the convex surface with positive curvature shrinks, and eventually disappears at time $t=t_f$, so that the net order parameter of the system approaches its equilibrium value $\phi_0$ as $\Delta \hat{\phi} \propto (t_f-t)^{(d^*+1)/2}$, where $d^*=d-1$ is the dimension of the interface, \textit{i.e.} $d^*=2$ for a spherical domain and $d^*=1$ for a cylindrical domain. 

Based on our observations of the growth of individual domains, we may quantify the global dynamics of phase amplification by considering the shrinking and growing of multiple domains. Consider the number of growing phase domains as $(\ell q)^d$, where $\ell$ is the linear size of the system, $q$ is the characteristic curvature (wave number) of the domain's interface, and $d$ is the dimensionality of the space in which the domain may grow. If we neglect the interaction between different domains, the root-mean-square (RMS) fluctuation of the number of shrinking or growing domains with positive or negative curvature per unit volume is $\sqrt{(\ell q)^d}/\ell^d$. To obtain the growth rate of the average order parameter, this factor must be multiplied by the growth rate ($\upsilon/t$) of the phase domain volume ($\upsilon \propto q^{-d}$). From our observations of Fig.~\ref{Fig-timedep}a, the size of the domains will grow as a square root of time, $t\propto q^{-2}$, then $\upsilon/t\propto 1/q^{d-2}$. Thus, the growth rate of the averaged absolute value of the order parameter is $\partial \langle\left|\phi\right|\rangle/\partial t \propto \ell^{-d/2}q^{-d/2+2}$, and as a function of time this is equivalent to $\partial \langle\left|\phi\right|\rangle/\partial t \propto \ell^{-d/2}t^{d/4-1}$. Integrating gives
\begin{equation}\label{Eq-phivst}
    \langle\left|\phi(t,\ell,T)\right|\rangle = A \phi_0(T)\left(\frac{t}{\tau}\right)^{d/4}
\end{equation}
where the characteristic time $\tau = \tau_0\ell^2$, with $\tau_0$ being inversely proportional to $p_r$ (see the inset of Fig. \ref{fig-scaling}b), and $\phi_0(T)$ is the equilibrium value of the order parameter. The amplitude, $A$, is practically independent of temperature. It can be shown from the Kelvin equation that $A \propto \sigma\xi/\phi_0^2$ \cite{Huang}, where $\sigma$ is the interfacial tension. This combination is a constant in mean-field theory in which $\sigma \propto \phi_0^3$ and $\xi \propto \phi_0^{-1}$, while in the Ornstein-Zernike approximation of scaling theory $\sigma \propto \xi^{-2}$ and $\phi_0^2\propto \xi^{-1}$ \cite{fisher_scaling_1983,anisimov_chapter_2010}. Thus, for all practical purposes, $A$ can be incorporated as a constant factor in $\tau$. Eq.~(\ref{Eq-phivst}), as demonstrated in Fig.~\ref{fig-scaling}a, is strongly supported by the simulation data. It is shown that the growth of the average order parameter, in $d=3$ space, closely follows the scaling law $\langle|\phi|\rangle \propto t^{3/4}$ for $t \ll \tau$ - see the inset of Fig.~\ref{fig-Averages}d - while, for $t \gg \tau$, it is constant. Therefore, Eq.~(\ref{Eq-phivst}) can be presented in the scaling form $\langle |\phi|\rangle/\phi_0(T) = f(X)$, where $X = t/\tau(\ell)$ and $f(X)$ is a scaling function such that $f(X)\to X^{d/4}$ for $X\ll 1$ and $f(X)=1$ for $X\gg 1$.

\paragraph*{Observations of Phase Amplification for Small $p_r$}- While systems simulated with relatively large Glauber probabilities ($p_r \geq 0.01$) all collapse to the same master curve, as shown in Fig.~\ref{fig-scaling}b, systems with relatively small probabilities ($p_r < 0.01$) have a larger average amplification rate (slope) at the inflection point. Therefore, for large $p_r$ the time evolution of the phase domain is controlled by fast, local interconversion due to Glauber dynamics, rather than the slow, global diffusion due to Kawasaki dynamics, which only dominates at very small Glauber-interconversion probabilities ($p_r < 0.01$). We attribute the dramatic change in the amplification rate, shown by the deviation from the master curve in Fig.~\ref{fig-scaling}b, to the increased chance of systems entering a metastable zero-curvature phase-domain state for smaller $p_r$. We clarify the effect of a system entering this metastable state in Fig.~\ref{Fig-timedep}b, by allowing the system to reach an equilibrium Schwarz-P interface under Kawasaki dynamics ($p_r=0$). The collapse of the Schwarz-P interface occurs after switching the system to {Glauber} dynamics ($p_r=1$) at $t = t_0$. When a zero curvature interface is first formed due to Kawasaki dynamics, then (no matter how small the Glauber probability, $p_r$) Glauber dynamics will proceed until, eventually, one of the phases will disappear. In this case, we use a renormalized time $t_G=t p_r$, which essentially counts only the Glauber steps. The sigmoidal shape of this curve shows a crossover from random, square root behavior (shown in the inset of Fig.~\ref{Fig-timedep}b) corresponding to the initial random walk of the interface near unstable equilibrium to the quick exponential amplification when a part of the interface develops non-zero curvature. This phenomenon explains the increase in the rate of the growth of the order parameter for the small probability of Glauber dynamics in Fig.~\ref{fig-scaling}b through the transformation of $t$ to $t_G = t/\tau_0 \propto p_r$.

\paragraph*{Conclusion} - In this work, we conceptualize the phenomenon of phase amplification, the growth of one phase at the expense of another one, and quantitatively characterize the speed of phase amplification through scaling laws based on the Kelvin equation. We simulate this phenomenon using a hybrid model, which combines Glauber-interconversion and Kawasaki-diffusion dynamics, in an Ising system. The developed approach is applicable to a broad spectrum of Ising-like systems with mixed dynamics. Such systems include ferromagnets, ferroelectrics, liquid  crystals, and materials with order-disorder transitions and chemical reactions. {For example, the developed approach explains the results of the first computational study of phase amplification in chiral crystals \cite{chiral_Debenedetti_2013}, and can also be used to describe chiral amplification in a mixture of interconverting enantiomers \cite{latinwo_molecular_2016} and real fluids (lacking the symmetry of the lattice gas  \cite{anisimov_nature_2006}) - if they exhibit interconversion}. In the future, we are interested in investigating phase amplification in polyamorphic liquids \cite{anisimov_thermodynamics_2018} and in systems with coupled order parameters \cite{takae_role_2020}.

\begin{acknowledgments}
We want to thank Pablo Debenedetti, Hans J. Herrmann, and Ivan Saika-Voivod for useful comments. This work is a part of the research collaboration between the University of Maryland, Princeton University, Boston University, and Arizona State University supported by the National Science Foundation. The research at the University of Maryland was supported by NSF award no. 1856479. The research at Boston University was supported by NSF award no. 1856496. SVB acknowledges the partial support of this research through the Dr. Bernard W. Gamson Computational Science Center at Yeshiva College.
\end{acknowledgments}

\bibliography{PhaseBullying}

\end{document}